\documentclass[12pt,a4paper]{article}
\usepackage[dvips]{epsfig}
\usepackage{multicol}

\title{Energy dependence of\\ 
transverse mass spectra of kaons\\ 
produced in p+p 
and p+$\overline{\rm{p}}$ interactions.\\ A compilation.}
\author{Michael Kliemant$^{a}$, Benjamin Lungwitz$^{a}$ \\
and Marek Ga\'{z}dzicki$^{a,b}$}
\date{\today}

\begin{document}
\maketitle


\noindent
\begin{minipage}[t]{12.5cm}

$^a$ Institut f\"ur  Kernphysik, Universit\"at  Frankfurt,
Germany\\
$^b$ \'Swi\c{e}tokrzyska Academy, Kielce, Poland \\
\end{minipage}

\vspace*{0.4cm}


\begin{abstract}
The data on $m_{T}$ spectra of 
$K^{0}_S$, $K^{+}$ and $K^{-}$ mesons 
produced in all inelastic p+p and p+$\overline{\rm{p}}$ 
interactions in the energy range $\sqrt{s_{NN}}=4.7-1800$~GeV
are compiled and analyzed.
The spectra are parameterized by a single exponential function, 
$\frac{dN}{m_{T}dm_{T}}=C\cdot e^{-m_{T}/T}$, 
and the inverse slope parameter $T$ is 
the main object of study.
The $T$  parameter is found to be similar 
for $K^{0}_{s}$, $K^{+}$ and $K^{-}$ mesons.
It increases monotonically with collision energy 
from $T\approx130$~MeV
at $\sqrt{s_{NN}}=4.7$~GeV to $T\approx220$~MeV at $\sqrt{s_{NN}}=1800$ GeV.
The $T$ parameter measured in p+p($\overline{\rm{p}}$) interactions
is significantly lower than the corresponding 
parameter obtained for central Pb+Pb collisions at all studied energies.
Also the shape of the energy dependence of $T$ is different for 
central Pb+Pb collisions and p+p($\overline{\rm{p}}$) interactions.
\end{abstract}

\section{Introduction}

The search for signals of the deconfinement phase transition and thus evidence for the existence 
of a quark gluon plasma in nature was the 
main objective of the experimental study of nucleus--nucleus (A+A) collisions at high energies.

Recent results obtained in broad range of energy allowed to identify anomalies in the energy dependence
of hadron production in central Pb+Pb
collisions \cite{Afanasiev:2002mx,Alt:2003rn}. 
They suggest that, in fact, the onset of deconfinement 
is observed and is located at the low CERN SPS energies
$\sqrt{s_{NN}} \approx 7.5$ GeV 
\cite{Gazdzicki:1998vd,Gorenstein:2003cu}.
Among basic observables used in this study are pion 
and kaon multiplicities and transverse 
mass spectra of kaons.
In the identification of the effects related to the onset 
of deconfinement in A+A collisions a 
comparison with the
corresponding results obtained in nucleon-nucleon (N+N) 
interactions plays a special role. 
In case of the study of energy dependence of pion and kaon 
multiplicities this comparison was done based on the existing compilations of the data on 
all inelastic proton-proton (p+p),
proton-neutron (p+n) and proton-antiproton (p+$\overline{\rm{p}}$) interactions 
\cite{Gazdzicki:zs,Gazdzicki:1996pk}.

A recently found \cite{Gorenstein:2003cu} anomaly in the energy 
dependence of the shape of transverse
mass spectra of kaons
produced in central Pb+Pb collisions raised already significant interest 
\cite{Mohanty:2003yn,Mohanty:2003fy,Bratkovskaya:2003ie,Gazdzicki:2003dx,Grassi:2003dc}.
However, the results on transverse mass spectra of kaons in N+N interactions were up to 
now not compiled and a detailed comparison with the corresponding A+A data was missing.
The aim of our paper is to fill this gap.\\
The paper is organized as follows.
In section~2 the existing data are reviewed and analyzed.
The energy dependence of the spectra is presented 
and discussed
in section~3. The paper is closed by the summary,
section~4.\\
					
\section{A compilation of p+p($\overline{\textbf{p}}$) data}

Most of the results on kaon transverse 
momentum ($p_T$) spectra in all inelastic
p+p($\overline{\rm{p}}$) interactions are obtained 
for $K^{0}_{S}$ mesons and we start our compilation from 
the review and analysis of these data.
Further on we compile and analyze data on charged kaon ($K^{+}$, $K^{-}$) production.

In most cases the original experimental papers present $p_T$ spectra in the form 
$\frac{d^{3}n}{dp^2_Tdy}$ or $\frac{Ed^3n}{d\vec{p}}$. 
From these results the transverse mass
$m_T$ 
($m_T=\sqrt{p^2_T+m^2_0}$, where $m_0$ is the particle rest mass)
spectra $\frac{d^{2}n}{m_{T}dm_{T}dy}$ can be easily obtained
($\frac{d^{2}n}{m_{T}dm_{T}dy}\sim\frac{d^{3}n}{dp^2_Tdy}\sim\frac{Ed^3n}{d\vec{p}}$).

Our compilation and analysis is limited to the low $p_T$ region, 
$p_T\leq1.3$~GeV/c.
First, it is because in this region the energy dependence of 
the $m_T$ spectrum of kaons
produced in A+A collisions was studied. 
Second, only in this region an exponential parameterization
of the p+p($\overline{\rm{p}}$) data as used for the analysis of A+A collisions
\cite{Afanasiev:2002mx,Alt:2003rn}
\begin{equation}
\label{eq1}
\\
\frac{dn}{m_{T}dm_{T}}=C\cdot e^{-m_{T}/T}
\\
\end{equation}
is approximately valid
\footnote{ 
At high $m_T$ the spectra obey a power law behavior, 
$\frac{dn}{m_{T}dm_{T}} \sim m_{T}^{-P}$,
which is interpreted as due to 
hard (parton) scattering.}.
In Eq.~\ref{eq1} the inverse slope parameter $T$ 
and the normalization parameter $C$
are treated as free parameters and their values are extracted from 
the least square fits to the experimental spectra.

The compiled $p_T$ spectra are measured either at midrapidity (y$\approx0$) 
or they are integrated 
over forward or backward hemispheres starting from midrapidity.
The rapidity of kaons peaks at midrapidity [21-29]
and thus the shape of the integrated spectra is 
dominated by the contribution from this region. 
In the following we do not distinguish between both types of measurements.

\subsection{$K^{0}_{S}$ spectra}

Transverse momentum spectra of $K^{0}_S$ mesons produced in
all inelastic p+p interactions [21-29] are measured 
in fixed target, mostly bubble chamber experiments, 
at energies below $\sqrt{s_{NN}}\approx30$~GeV. 
At higher energies, $\sqrt{s_{NN}}= 200 - 900$~GeV, the measurements  
for
p+$\overline{\rm{p}}$ interactions are performed at the Sp$\overline{\rm{p}}$S collider
\cite{Ansorge:fq,Bocquet:1995jq}.
The $K^{0}_{S}$ data come from the analysis of charged decays of 
$K^{0}_{S}$ mesons,
$K^{0}_{S}$~$\rightarrow\pi^{+}+\pi^{-}$.
A characteristic feature of these results are relatively 
low systematic uncertainties and large 
statistical errors due to the small number of analyzed events.

The summary of the data on $m_{T}$ spectra of $K^{0}_{S}$ 
mesons is given in Tables 1 and 2, where
$\sqrt{s_{NN}}$, the 
$p_{T}$ range selected for the analysis, the rapidity~(y) range in which
the measurement was done and the 
reference to the original experimental papers are given. 
The $m_T$ spectra are plotted as a
function of $m_T-m_0$ in Fig.~1.
The normalization of the spectra is arbitrary. They are ordered from
bottom to top according to increasing energy.

The spectra displayed in Fig.~1 are fitted by an exponential function, Eq.~\ref{eq1},
in the whole $m_T$ range ($m_T-m_0\leq0.85$~GeV/c$^2$) as well as 
in two subintervals, the ``low--$m_T$'' interval ($m_T-m_0\leq0.25$~GeV/c$^2$) and
the ``high--$m_T$'' interval ($0.25 < m_T-m_0 \leq0.85$~GeV/c$^2$). 
The inverse slope parameter $T$ and $\chi^{2}/NDF$ resulting from the fits
in the whole interval are given in Tables~1 and 2. 
The corresponding functions are plotted in Fig.~1 by solid lines.
It is seen that the used parametrization (Eq.~\ref{eq1}), 
reasonably well describes $K^{0}_{S}$ spectra in the 
whole $m_T$ range  at all studied energies 
($\sqrt{s_{NN}}=4.7-900$~GeV), both
for p+p and p+$\overline{\rm{p}}$ interactions.
Differences observed between results obtained for
``low--$m_T$'' and ``high--$m_T$'' intervals are shown and  
discussed in section~3.

\subsection{$K^{+}$ and $K^{-}$ spectra}

A summary of the data used in the analysis of the $m_{T}$ spectra 
of $K^{+}$ and $K^{-}$ mesons produced
in p+p interactions at the ISR \cite{Alper:1975jm} 
and fixed target experiments \cite{Kraus:2003ij,Zabrodin:ca}
is given in Tables 3 and 4. 
In Table 5 the data on 
charged kaon spectra in p+$\overline{\rm{p}}$ interactions 
at the Sp$\overline{\rm{p}}$S \cite{Banner:jq}
and the Tevatron \cite{Alexopoulos:hn,Alexopoulos:wt} are compiled. 
In most cases the $K^+$ and $K^-$ identification is
done by means of time of flight \cite{Alexopoulos:hn,Alper:1975jm} and 
energy loss measurements  \cite{Kraus:2003ij} 
or by use of
gas Cerenkov counters \cite{Alper:1975jm}.
In the bubble chamber
experiment \cite{Zabrodin:ca} the kaons were identified by the analysis of their decays.

The $m_{T}$ spectra of $K^{+}$ and $K^{-}$ mesons are plotted in Figs.~2 and 3 together with the 
exponential functions (Eq.~\ref{eq1})
fitted in the whole $m_T$ interval.
The corresponding $T$ parameter and $\chi^{2}/NDF$ are given in Tables 3, 4 and 5.
Majority of the $K^{+}$ and $K^{-}$ spectra follow reasonably well
the parametrization. 
The most significant deviations are observed
for the $K^+$ spectra at $\sqrt{s_{NN}}=63$~GeV and the $K^-$ spectra
at $\sqrt{s_{NN}}=53$ and $63$ GeV. 
The results obtained for the fits in
``low--$m_T$'' and ``high--$m_T$'' intervals, which allow to quantify the observed
deviations, are shown and 
discussed in section~3.

\section{The energy dependence}

\subsection{Results}
The  energy dependence of   
the $T$ parameter fitted in the whole $m_T$ interval 
($m_T-m_0\leq0.85$~GeV/c$^2$)
to the $m_T$ spectra of
kaons  produced  
in all inelastic  p+p($\overline{\rm{p}}$) interactions
is shown in Fig.~4.
No significant differences are seen between results
obtained for $K^{0}_S$, $K^{+}$ and $K^{-}$ mesons 
as well as between data for p+p and p+$\overline{\rm{p}}$
interactions. 

The dependence of the $T$ parameter on $\sqrt{s_{NN}}$ was fitted by an expression:
\begin{equation}
\label{eq2}
\\
T=a+b\cdot\ln{\sqrt{s_{NN}}},
\\
\end{equation}
where $a$ and $b$ are fit parameters 
and $\sqrt{s_{NN}}$ is given in units of GeV.
The best fit to the data presented in Fig.~4 yields $a=(115\pm2.8)$~MeV, $b=(13.7\pm0.7)$~MeV 
and $\chi^{2}/NDF=120/30$.
For most of the points the observed spread around the parameterization is consistent with one 
expected from statistical fluctuations. Only the points at $\sqrt{s_{NN}}=23.7$~GeV \cite{Lopinto:ct},
$\sqrt{s_{NN}}=540$~GeV \cite{Alexopoulos:wt} and $\sqrt{s_{NN}}=900$~GeV \cite{Ansorge:fq}
deviate by more than 3 standard deviations.
The compiled results indicate that $T$ increases monotonically with collision energy from
$T\approx130$~MeV at $\sqrt{s_{NN}}=4.7$ to $T\approx220$~MeV at $1800$~GeV.

The inverse slope parameters obtained form the fits performed
in the ``low--$m_T$'' and ``high--$m_T$'' intervals are plotted as 
a function of collision energy in Fig.~5. 
Up to $\sqrt{s_{NN}} \approx 20$ GeV the $T$ values for 
the ``low--$m_T$'' and ``high--$m_T$'' intervals are approximately equal.
At higher energies the $T$ parameter for ``low--$m_T$'' interval is systematically
lower than for the ``high--$m_T$'' interval. 
For the ``low--$m_T$'' interval the $T$ parameter is 
approximately independent of the collision energy, 
whereas in the ``high--$m_T$'' interval it
seems to increase. 
The observed behaviour may  be attributed to  
rapid development of the power law 
tail with
increasing collision energy.

The data on the $T$ parameter 
were fitted by the logarithmic function, Eq.~\ref{eq2}, for each interval separately.
The resulting parameters 
and $\chi^{2}/NDF$ are:
$a=(139\pm7.7)$~MeV, $b=(-0.1\pm2.2)$~MeV, $\chi^{2}/NDF=39/24$ and
$a=(110\pm7)$~MeV, $b=(18.2\pm1.7)$~MeV, $\chi^{2}/NDF=60/24$ for
``low--$m_T$'' and ``high--$m_T$'' intervals, respectively. 
The functions fitted in both subintervals and the whole $m_T$
interval are shown in Fig.~5 for a comparison.

In Figs.~6 and 7 the results on the $T$ parameter of 
kaons obtained for p+p($\overline{\rm{p}}$) 
interactions are plotted  as a function of $\sqrt{s_{NN}}$ 
together with the corresponding data for central Pb+Pb (Au+Au)
collisions at AGS \cite{Ahle:2000wq}, 
SPS \cite{Afanasiev:2002mx,Alt:2003rn} 
and RHIC \cite{Ouerdane:2002gm,Adler:2002wn}.
The data points for central Pb+Pb (Au+Au) collisions and p+p($\overline{\rm{p}}$)
interactions were extracted from the  fits performed in the whole $m_T$ interval.
A different dependence is measured in Pb+Pb (Au+Au) collisions than in 
p+p($\overline{\rm{p}})$ interactions.
At all energies the $T$ parameter is 
significantly higher in Pb+Pb (Au+Au) collisions than
in elementary interactions. 
In the case of heavy ion collisions  a well developed 
 plateau at SPS energies is observed.
There is no obvious sign of such behaviour for 
p+p($\overline{\rm{p}})$) interactions, 
however, the precision of these data 
is low and they can not exclude anomalies similar to this observed for
Pb+Pb (Au+Au) collisions.

\subsection{Discussion}
The compilation of the p+p($\overline{\rm{p}}$) data performed  in this paper
was triggered by the observation of the anomalous energy dependence of the
shape of transverse mass spectra of kaons measured in central Pb+Pb (Au+Au) 
collisions and its possible relation to the deconfinement phase transition 
\cite{Gorenstein:2003cu}.
Within statistical--hydrodynamic approach to A+A collisions the 
heavy ion data shown in Fig.~6 are 
interpreted as follows.

The $T$ parameter increases strongly with collision energy up to
the lowest CERN SPS energies ($\sqrt{s_{NN}} \approx 7$ GeV). 
This is an energy
region where the creation of confined matter at the early stage of the
collisions is expected. Increasing collision energy leads to an increase
of the early stage temperature and pressure. This results in grow of 
the transverse collective flow velocity, $\overline{v}_{T}$, and temperature 
$T_{f}$, at freeze--out. 
Consequently the transverse activity of produced hadrons, measured by the
inverse slope parameter, increases with increasing energy. 

The $T$ parameter is approximately independent of the collision energy
in the SPS energy range, $\sqrt{s_{NN}} = 7 - 20$ GeV. 
In this energy
region the transition between confined and deconfined matter is expected
to be located. The resulting modification of the equation of state 
``suppresses'' an increase of both $T_{f}$ and $\overline{v}_{T}$, and
this leads to the observed plateau structure in the energy dependence of
the $T$ parameter. 

At RHIC 
($\sqrt{s_{NN}}=130$~GeV and 200~GeV) the $T$
is significantly larger than at SPS energies. 
In the RHIC energy
domain the equation of state at the early stage becomes again stiff, the
early stage temperature and pressure increase with collision energy. This
results in increase of the transverse flow $\overline{v}_{T}$ and,
consequently, in increase of $T$  between SPS and RHIC
energies. 

Within the above picture there are two basic differences between central
collisions of heavy nuclei and p+p interactions.

First, collective expansion effects are expected to be important only
in heavy ion collisions, as they result from the pressure gradient
generated in the dense interacting matter. 
The expansion leads to an increase of the $T$ parameter and consequently
one predicts a lower values of $T$ in p+p interactions,
where collective expansion is absent and hadrons are emitted
directly from the fragmentation of strings, than in central A+A collisions.
This qualitative prediction is independent of the state of matter created
at the early stage of the collisions and thus it is valid at all collision
energies.
In fact, it is confirmed by the data compiled in this paper  (see Fig.~6).  

Second, the stationary value of $T$ at SPS energies is related to the modification
of the equation of state of matter created at the early stage for mixed phase 
and thus relative suppression of the collective expansion.
As the collective expansion is absent in p+p interactions and, in general,
the applicability of statistical models is in question for elementary
interactions, one does not expect to observe any anomaly in energy dependence
of $T$ in p+p interactions.
This prediction is not in contradiction with the data compiled in this paper.
However, large experimental errors  allow to exclude only relatively large
effects.  

The string hadronic models (Fritiof \cite{Andersson:1992iq}, 
UrQMD \cite{Bleicher:1999xi}, HSD \cite{Geiss:1998ki, Cassing:es}) well 
describe results on $T$  for p+p interactions \cite{Bratkovskaya:2003ie}. 
Models  without rescattering (see e.g. Fritiof \cite{Andersson:1992iq}) predict  similar
values of $T$ for p+p interactions and central A+A collisions.
These approaches are obviously excluded by the data.
The models with rescattering of initially produced strings and
hadrons should, in general, lead to an increase of $T$ for central A+A collisions.
Surprisingly, it was recently found \cite{Bratkovskaya:2003ie} within HSD and UrQMD models, that the
rescattering only weakly influences shape of the transverse mass spectra
of kaons.
Thus also these approaches significantly under-predict the $T$ parameter measured
in central A+A collisions at all energies \cite{Bratkovskaya:2003ie}.

Within microscopic picture of A+A collisions one may try to speculate whether
a part of the observed increase of $T$ when going from p+p interactions to 
central A+A collisions is related to the higher excitation of  
nucleons in A+A collisions than in all inelastic  
interactions.
The experimental results, which can shed light on this subject are
poor, but nevertheless, at collision energies up to top SPS
energy they seem to contradict this hypothesis.
Relatively rich data on pion production in p+p interactions as
a function of multiplicity of produced hadrons show that
the mean transverse momentum (and the $T$ parameter) 
decreases with increasing multiplicity  \cite{Golokhvastov:2002cf}. 
Independence of the shape of $p_T$ spectrum  of kaons 
on multiplicity is suggested
by the p+p results at 12-200 GeV 
\cite{Jaeger:1974in}.
One should note, however, that
at Tevatron energies an increase of $\langle p_T \rangle$ of
pions and kaons with multiplicity is observed \cite{Alexopoulos:hn}.

\section{Summary}

We compiled and analyzed data on $m_{T}$ spectra  of 
$K^{0}_S$, $K^{+}$ and $K^{-}$ mesons 
produced in all inelastic p+p and p+$\overline{\rm{p}}$ interactions.
The spectra can be reasonably well described by a simple exponential parametrization 
$\frac{dN}{m_{T}dm_{T}}=C\cdot e^{-m_{T}/T}$
in the whole analyzed $m_T$ interval ($m_{T}-m_{0}\leq$ 0.85 $GeV/c^{2}$).
The values of the parameter $T$ measured for $K^{0}_{S}$, $K^{+}$ and $K^{-}$ are similar.
We do not observe any significant difference between data from p+p and p+$\overline{\rm{p}}$ interactions.
The $T$ parameter 
increases monotonically with $\sqrt{s_{NN}}$ from $T\approx130$~MeV
at $\sqrt{s_{NN}}=4.7$~GeV to $T\approx220$ MeV at $\sqrt{s_{NN}}\approx1800$~GeV.
This dependence can be parametrized as $T=(115\pm2.9)$~MeV~+~$(13.7\pm0.7)$~MeV~$\cdot\ln(\sqrt{s_{NN}})$,
where $\sqrt{s_{NN}}$ is given in units of GeV.
The $T$ measured in p+p ($\overline{\rm{p}}$) is significantly lower than the corresponding 
measurement in Pb+Pb (Au+Au) collisions.
Also the shape of the energy dependence of $T$  on energy is
different for central  Pb+Pb collisions and  p+p($\overline{\rm{p}}$) interactions. 

\vspace{0.3cm}

\noindent {\bf Acknowledgments.} We thank Rainer Renfordt for reading and
commenting our work and our colleagues at IKF and Marco van Leeuwen (NIKHEF) for discussions and
comments. Partial support by Bundesministerium f\"ur Bildung und Forschung
and by Polish Committee of Scientific Research under grant 2P03B04123
(M.~G.) is acknowledged.


\pagebreak

\begin{table}[p]
\caption{Summary of the data on the $p_{T}$ spectra of $K^{0}_{S}$ produced
in p+p interactions. The collision energy $\sqrt{s_{NN}}$, the $p_{T}$-range used for the
analysis, the c. m. rapidity range 
in which $p_T$ spectrum was measured
and the references to the original papers are given.
The fitted inverse slope parameter $T$ and $\chi^{2}/NDF$ are
also presented.}
~\\[0.1cm]
\begin{tabular}{|c|c|c|c|c|c|}
\hline
&&&&&\cr
$\sqrt{s}$(GeV) & $p_{T}$-range(GeV/c) & y-range & $T$(MeV) & $\chi^{2}/NDF$ & ref. \cr
&&&&&\cr
\hline

4.74 & 0-1.025 & 0-2.0 & 130.2 $\pm$ 3.3 & 8.2/6 & \cite{Blobel:1973jc} \cr
4.82 & 0-0.906 & 0-1.7 & 135.5 $\pm$ 11.4 & 0.45/4 & \cite{Jaeger:1974in} \cr
6.7 & 0-1.025 & 0-2.0 & 141.2 $\pm$ 3.4 & 14/6 & \cite{Blobel:1973jc} \cr
11.4 & 0-0.945 & 0-2.4 & 148.3 $\pm$ 10.5 & 2.4/8 & \cite{Ammosov:1975bt} \cr
13.8 & 0-0.951 & 0-2.0 & 168.3 $\pm$ 17.5 & 1.4/4 & \cite{Chapman:fn} \cr
16.6 & 0-0.897 & 0-0.6 & 149.3 $\pm$ 10.6 & 18/4 & \cite{Brick:vj} \cr
19.6 & 0-0.889 & 0-3.0 & 153.4 $\pm$ 10.7 & 1.2/6 & \cite{Jaeger:1974in} \cr
23.7 & 0-1.225 & 0-3.0 & 191.4 $\pm$ 4.8 & 17/6 & \cite{Lopinto:ct} \cr
23.7 & 0-0.787 & -3.0-0 & 156 $\pm$ 14.3 & 2.8/5 & \cite{Sheng:1974zn} \cr
25.9 & 0-1.161 & 0-3.2 & 152.7 $\pm$ 5.9 & 6.1/8 & \cite{Asai:1984dv} \cr
27.5 & 0-1.165 & 0-3.0 & 169.4 $\pm$ 8.1 & 4.3/7 & \cite{Kichimi:1979te} \cr
\hline
\end{tabular}
\end{table}

\begin{table}[p]
\caption{Summary of the data on the $p_{T}$ spectra of $K^{0}_{S}$ produced
in p+$\overline{\rm{p}}$ interactions. For details see the caption of Table 1.}
~\\[0.1cm]
\begin{tabular}{|c|c|c|c|c|c|}
\hline
&&&&&\cr
$\sqrt{s}$(GeV) & $p_{T}$-range(GeV/c) & y-range & $T$(MeV) & $\chi^{2}/NDF$ & ref. \cr
&&&&&\cr
\hline
200 & 0.41-1.16 & 0-3.5 & 183.9 $\pm$ 23.2 & 2.3/3 & \cite{Ansorge:fq}\cr
630 & 0.275-1.22 & $|\eta|<2.5$ & 208 $\pm$ 6.6 & 35/11 & \cite{Bocquet:1995jq}\cr
900 & 0.38-1.17 & 0-3.5 & 248.6 $\pm$ 12.4 & 4.2/4 & \cite{Ansorge:fq}\cr
\hline
\end{tabular}
\end{table}

\begin{table}[p]
\caption{Summary of the data on the $p_{T}$ spectra of $K^+$ produced
in p+p interactions. For details see the caption of Table 1.}
~\\[0.1cm]
\begin{tabular}{|c|c|c|c|c|c|}
\hline
&&&&&\cr
$\sqrt{s}$(GeV) & $p_{T}$-range(GeV/c) & y-range & $T$(MeV) & $\chi^{2}/NDF$ & ref. \cr
&&&&&\cr
\hline
17.2 & 0-1.3 & y$\approx$0 & 172 $\pm$ 17 & &  \cite{Kraus:2003ij}\cr
23 & 0.1-1.2 & y$\approx$0 & 161.5 $\pm$ 4.9 & 9.5/5 & \cite{Alper:1975jm} \cr
31 & 0.1-1.2 & y$\approx$0 & 169.5 $\pm$ 5.2 & 8.3/5 & \cite{Alper:1975jm} \cr
45 & 0.1-1.2 & y$\approx$0 & 155 $\pm$ 4.9 & 5/5 & \cite{Alper:1975jm} \cr
53 & 0.1-1.2 & y$\approx$0 & 167.5 $\pm$ 5.2 & 23/5 & \cite{Alper:1975jm} \cr
63 & 0.1-1.2 & y$\approx$0 & 179.4 $\pm$ 13.8 & 34/6 & \cite{Alper:1975jm} \cr

\hline
\end{tabular}
\end{table}

\begin{table}[p]
\caption{Summary of the data on the $p_{T}$ spectra of $K^-$ produced
in p+p interactions. For details see the caption of Table 1.}
~\\[0.1cm]
\begin{tabular}{|c|c|c|c|c|c|}
\hline
&&&&&\cr
$\sqrt{s}$(GeV) & $p_{T}$-range(GeV/c) & y-range & $T$(MeV) & $\chi^{2}/NDF$ & ref. \cr
&&&&&\cr
\hline
7.74 & 0-1.025 & y$\approx$0 & 146.9 $\pm$ 41.1 & 1.81/8 &  \cite{Zabrodin:ca} \cr
17.2 & 0-1.3 & y$\approx$0 & 164 $\pm$ 16 & &  \cite{Kraus:2003ij}\cr
23 & 0.1-1.2 & y$\approx$0 & 150.7 $\pm$ 5.1 & 6.7/5 & \cite{Alper:1975jm} \cr
31 & 0.1-1.2 & y$\approx$0 & 145.8 $\pm$ 5.8 & 11/5 & \cite{Alper:1975jm} \cr
45 & 0.1-1.2 & y$\approx$0 & 161.8 $\pm$ 6.2 & 5.9/5 & \cite{Alper:1975jm} \cr
53 & 0.1-1.2 & y$\approx$0 & 177.8 $\pm$ 5.6 & 33/5 & \cite{Alper:1975jm} \cr
63 & 0.1-1.2 & y$\approx$0 & 185.6 $\pm$ 13.9 & 31/6 & \cite{Alper:1975jm} \cr

\hline
\end{tabular}
\end{table}

\begin{table}[p]
\caption{Summary of the data on the $p_{T}$ spectra of $K^+$+$K^-$ produced
in p+$\overline{\rm{p}}$ interactions. For details see the caption of Table 1.}
~\\[0.1cm]
\begin{tabular}{|c|c|c|c|c|c|}
\hline
&&&&&\cr
$\sqrt{s}$(GeV) & $p_{T}$-range(GeV/c) & y-range & $T$(MeV) & $\chi^{2}/NDF$ & ref. \cr
&&&&&\cr
\hline
300 & 0.28-1.22 & y$\approx$0 & 192.6 $\pm$ 9.8 & 15.7/7 & \cite{Alexopoulos:wt} \cr
540 & 0.28-1.22 & y$\approx$0 &  158.9 $\pm$ 8.1 & 11.2/7 & \cite{Alexopoulos:wt} \cr
540 & 0.45-1.05 & y$\approx$0 & 215.7 $\pm$ 18.4 & 6.8/5 & \cite{Banner:jq} \cr
1000 & 0.28-1.22 & y$\approx$0 & 217.8 $\pm$ 6.7 & 11.2/7 & \cite{Alexopoulos:wt} \cr
1800 & 0.28-1.17 & y$\approx$0 & 213.1 $\pm$ 5.1 & 11/17 & \cite{Alexopoulos:hn} \cr

\hline
\end{tabular}
\end{table}
\newpage

\begin{figure}[p]
\label{K0sSum}
\epsfig{file=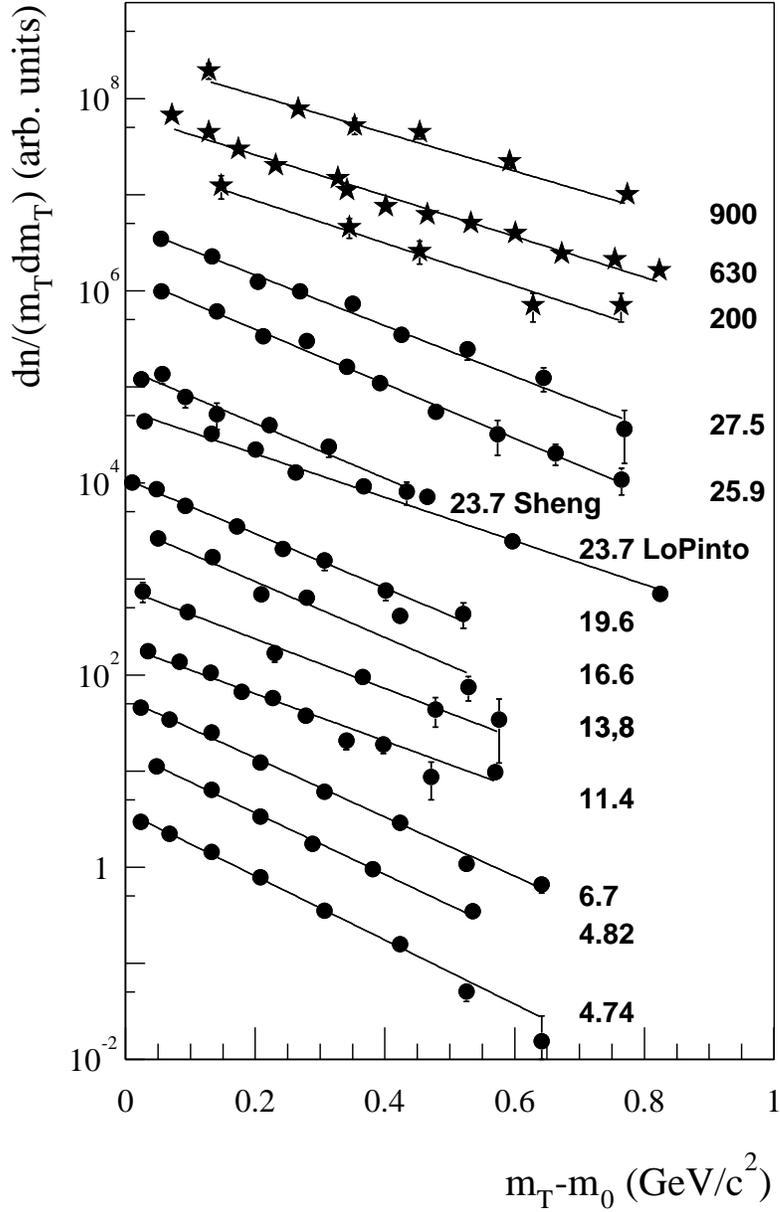,width=11.5cm}
\caption{Transverse mass spectra of $K^0_S$ mesons produced in p+p (circles) and p+$\overline{\rm{p}}$
(stars) interactions. The normalization of the spectra is arbitrary and 
the numbers next  to the
spectra give c. m. collision energy in GeV. The fits of the exponential function (Eq.~\ref{eq1}) are indicated by the solid lines.}
\end{figure}

\begin{figure}[p]
\label{K+Sum}
\epsfig{file=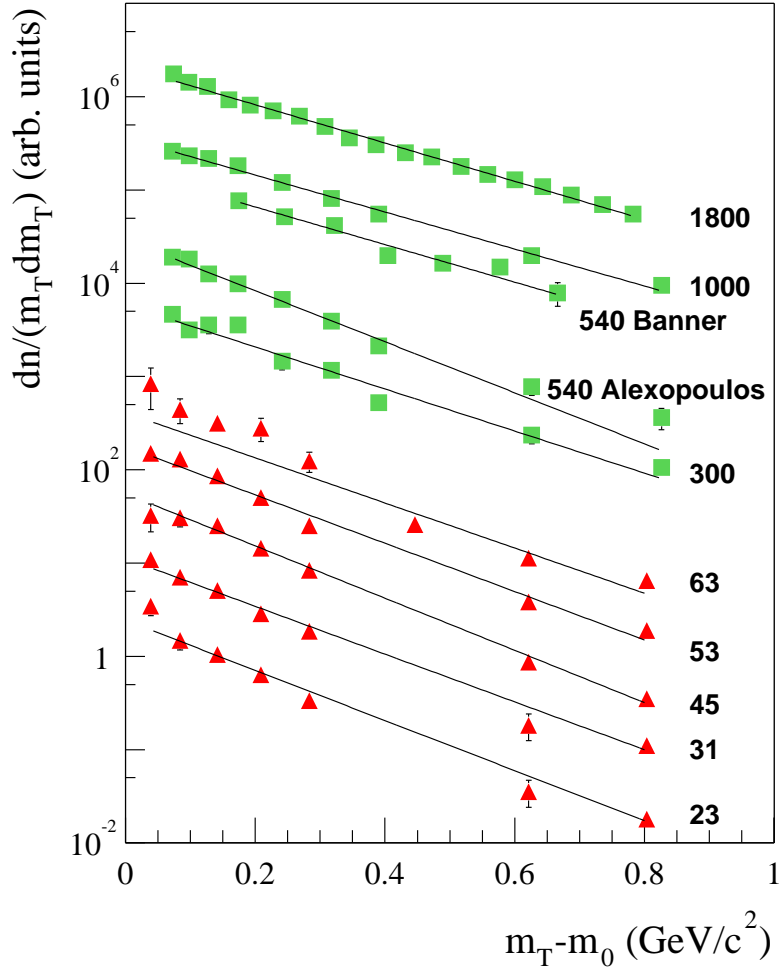,width=11.5cm}
\caption{(Color online) Transverse mass spectra of $K^+$ mesons produced in p+p interactions (triangles)
and $K^+$+$K^-$- mesons produced in p+$\overline{\rm{p}}$ interactions (squares). For details see the caption of Fig.~1.}
\end{figure}

\begin{figure}[p]
\label{K-Sum}
\epsfig{file=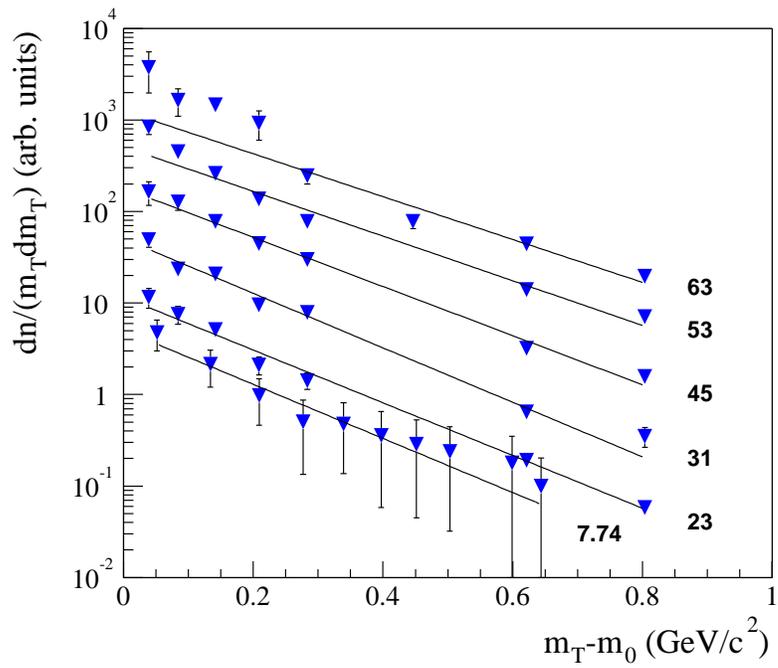,width=11.5cm}
\caption{(Color online) Transverse mass spectra of $K^-$ mesons produced in p+p interactions.
For details see the caption of Fig.~1.}
\end{figure}

\begin{figure}[p]
\epsfig{file=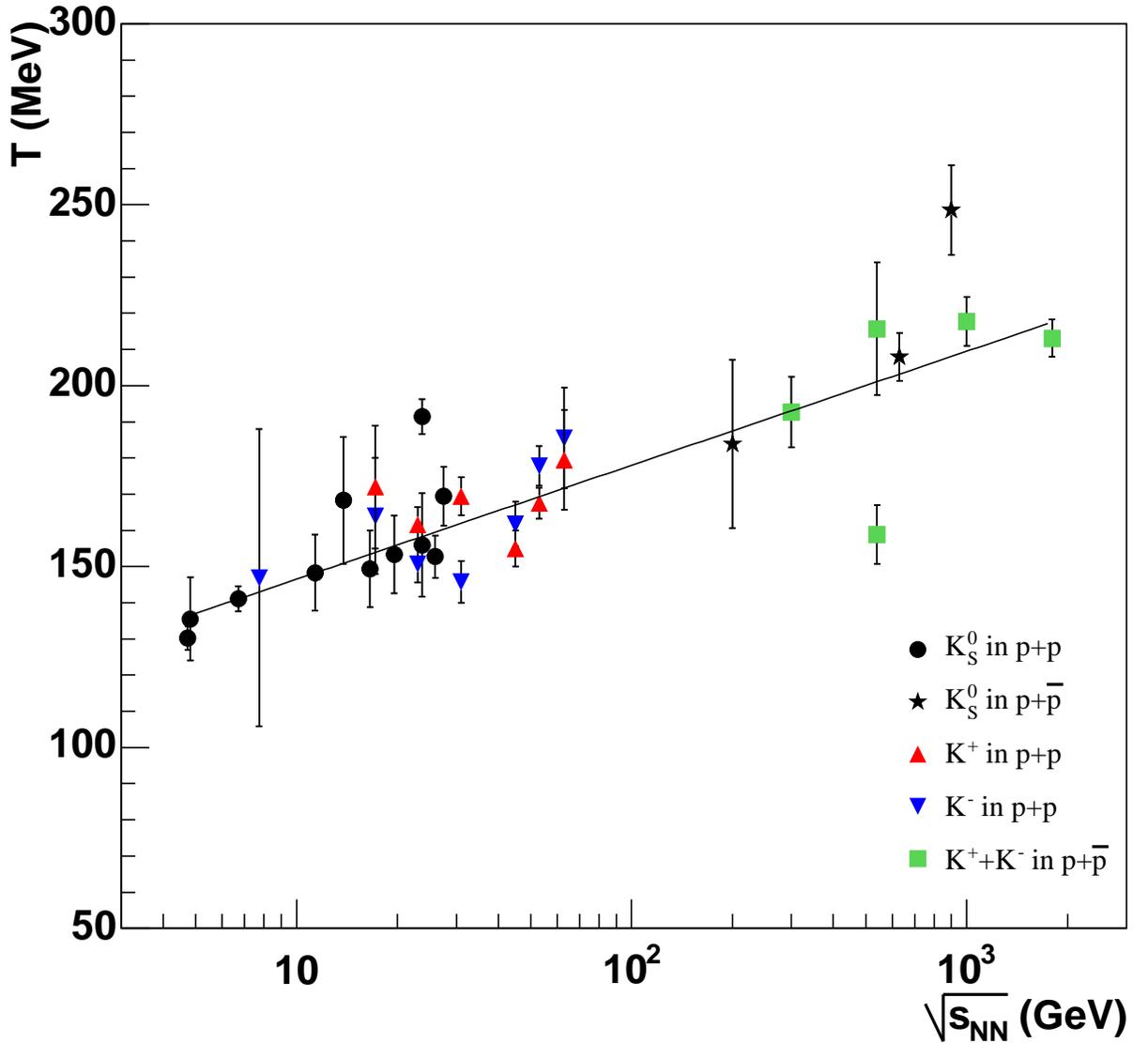,width=16cm,angle=0}
\caption{(Color online) Energy dependence of 
the inverse slope parameter $T$ of transverse mass 
spectra of $K^0_S$, $K^+$ and $K^-$ mesons produced
in p+p and p+$\overline{\rm{p}}$ interactions. 
The $T$ parameter was determined by fitting the spectra
(Eq. 1) in the whole analyzed $m_T$ interval,
$m_T -m_0 < 0.25$ GeV/c$^2$.
The logarithmic parameterization is indicated by solid line.}
\end{figure}

\begin{figure}[p]
\epsfig{file=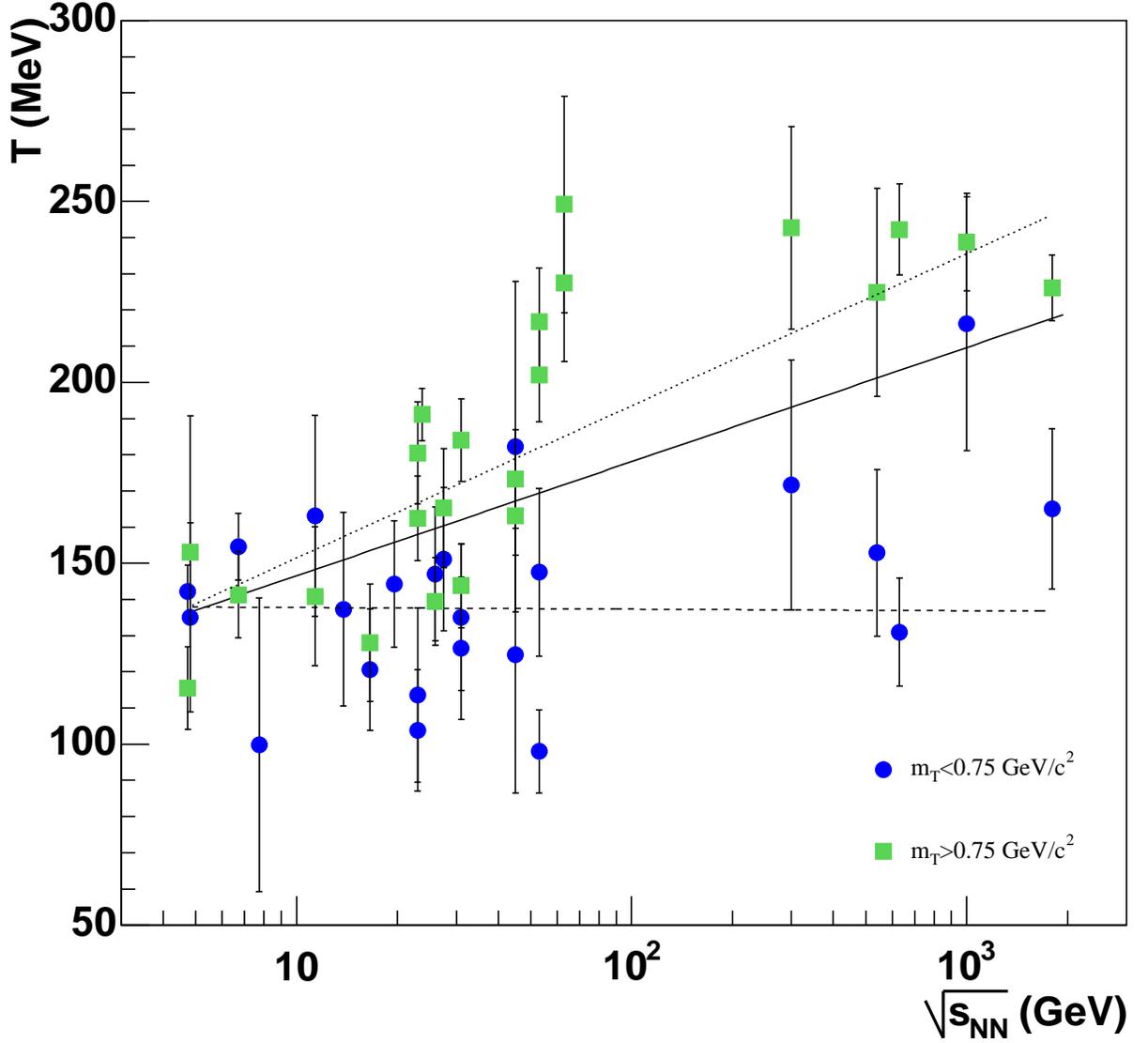,width=16cm}
\caption{(Color online) Energy dependence of the inverse slope parameter $T$ of transverse mass spectra of $K^0_S$, $K^+$ and $K^-$ mesons produced
in p+p and p+$\overline{\rm{p}}$ interactions, fitted separately in the ``high'' and ``low'' $m_T$ intervals.
The dashed and dotted lines show parameterizations of data for ``low'' and ``high'' $m_T$ intervals, respectively.
The parameterization obtained for the whole $m_T$ interval is indicated by a solid line for comparison.
Points with errors larger than 50 MeV are not plotted.}
\end{figure}

\begin{figure}[p]
\epsfig{file=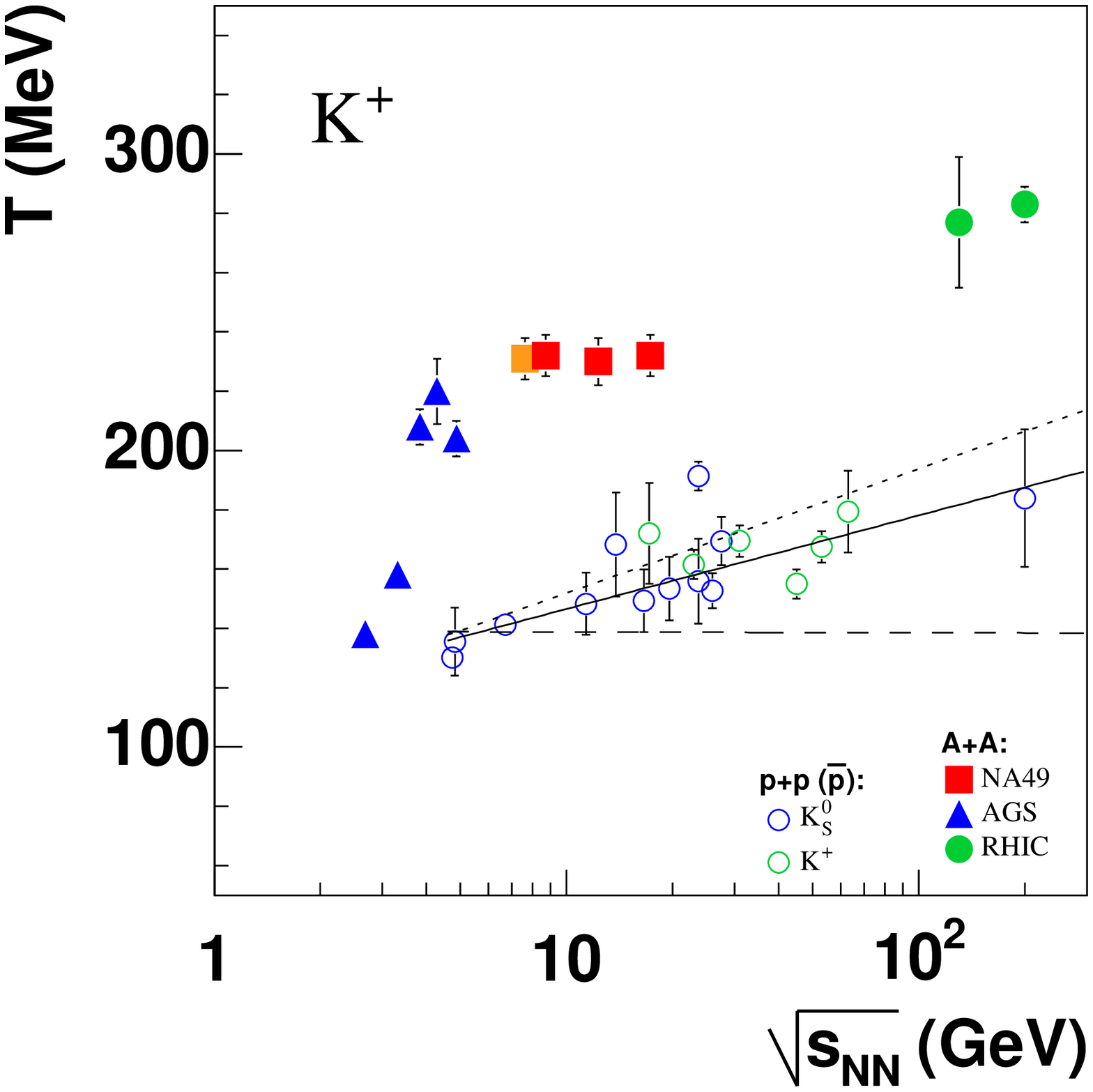,width=16cm}
\caption{(Color online) Energy dependence of the inverse slope parameter $T$ of transverse mass spectra of $K^+$ 
mesons produced in central collisions
of heavy nuclei (Pb+Pb \cite{Afanasiev:2002mx,Alt:2003rn}, Au+Au \cite{Ahle:2000wq,Ouerdane:2002gm,Adler:2002wn}) as well as $K^+$
and $K^0_S$ mesons produced in p+p and p+$\overline{\rm{p}}$ interactions. The solid, dashed and dotted lines indicate
parameterizations obtained for whole, ``low'' and ``high'' $m_T$ intervals, respectively.}
\end{figure}

\begin{figure}[p]
\epsfig{file=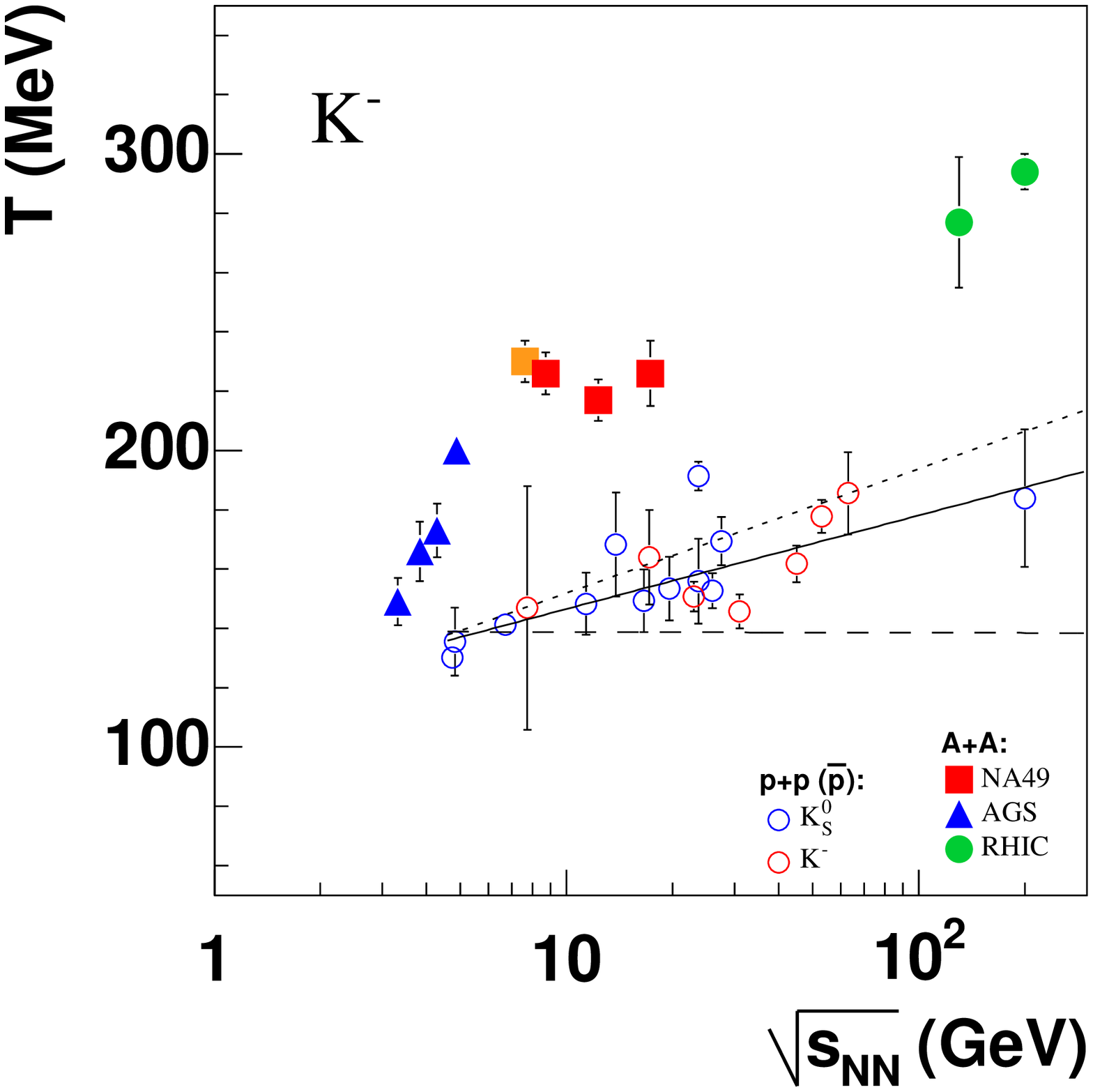,width=16cm}
\caption{(Color online) Energy dependence of the inverse slope parameter $T$ of transverse mass spectra of $K^-$ 
mesons produced in central collisions
of heavy nuclei (Pb+Pb \cite{Afanasiev:2002mx,Alt:2003rn}, Au+Au \cite{Ahle:2000wq,Ouerdane:2002gm,Adler:2002wn}) as well as $K^-$
and $K^0_S$ mesons produced in p+p and p+$\overline{\rm{p}}$ interactions. The solid, dashed and dotted lines indicate
parameterizations obtained for whole, ``low'' and ``high'' $m_T$ intervals, respectively.}
\end{figure}

\end{document}